\documentstyle[aps,prd,epsf]{revtex}

\newcommand{\hal}[1]{{#1 \over 2}}
\newcommand{\ket}[1]{| #1 \rangle}
\newcommand{\bra}[1]{\langle #1 |}

\newcommand{\delslash}{\partial \hspace{-6pt}/}
\def\Journal#1#2#3#4{{#1} {\bf #2}, #3 (#4)}

\def\NPA{{\em Nucl. Phys.} A}

\def\PLB{{\em Phys. Lett.}  B}
\def\PRL{{\em Phys. Rev. Lett.}}
\def\PRD{{\em Phys. Rev.} D}

\begin{document}
\draft
\preprint{TIT-HEP nnn/NP, hep-ph/9805306}
\title{Chiral Symmetry for Positive and Negative Parity Nucleons}
\author{D. Jido\thanks{jido@th.phys.titech.ac.jp},
       Y. Nemoto and M. Oka}
\address{Department of Physics, Tokyo Institute of
       Technology Meguro, Tokyo 1528551  Japan}
       \author{A. Hosaka}
\address{Numazu College of Technology
      3600 Ooka, Numazu 4108501 Japan}
\date{\today}
\maketitle

\begin{abstract}
   Chiral properties of positive and negative parity nucleons, $N$
   and $N^*$, are studied from the viewpoint of chiral symmetry.  Two
   possible ways to assign chiral transformations to the negative
   parity nucleon are considered.  Using linear sigma models
   based on the two chiral realizations, theoretical as well as
   phenomenological consequences of the two different assignments are
   investigated.  We find that the nucleon mass in the chiral restored
   phase is the key quantity to determine the meson-nucleon couplings 
   and the axial charges of nucleons.  We also discuss the role of chiral
   symmetry breaking in the mass splitting of
   $N$ and $N^*$ in the two sigma models.

\end{abstract}


\vspace{0.5cm}

Chiral symmetry and its spontaneous breakdown are very important to
understand properties of hadrons at low energy.  While the spectrum of
observed hadrons does not respect chiral symmetry, the broken chiral
symmetry is believed to be restored at finite temperature and/or at
finite density.  In the chiral restored phase, we expect that hadrons
in the same irreducible representation of the chiral group
are degenerate.  Hadrons in such a multiplet are called chiral partners
to each other.  So far the role of the chiral symmetry has been
extensively worked out for meson properties~\cite{hatsuda}.  For
instance, $(\sigma, \pi)$ and $(\rho, a_1)$ are candidates of chiral
partners, as they belong to the $(\hal{1},\hal{1})$ and $(1,1)$
representations of the chiral $SU(2)_L \times SU(2)_R$ group,
respectively.  In contrast, chiral symmetry of baryons is less
understood.  Chiral properties of hadrons are very important in
the study of, for instance, QCD phase transitions, since transition
properties depend crucially on the particle spectrum before and after
the transition.

So far there are not many works investigating negative parity nucleons
from the point of view of chiral symmetry~\cite{lee,coji}.
DeTar and Kunihiro (DK) studied positive and negative
parity nucleons in an extended $SU(2)$ linear sigma model~\cite{dk}.
Their work was motivated by the lattice QCD observations which
indicated existence of finite mass nucleons after the chiral symmetry
is restored~\cite{lattice}.  A similar observation was also made by 
Sch\"afer and Shuryak using an instanton liquid model~\cite{inst}.
When chiral symmetry is restored, one
would naively expect that nucleons become massless as is indicated in
the linear sigma model.  However, they showed that in the presence of
positive and negative parity nucleons it is possible to construct a
theory which allows finite nucleon masses without destroying chiral
symmetry.  Some of nucleon properties then depend crucially on the way
how chiral symmetry is implemented.

Recently we have studied properties of negative parity baryons using
the QCD sum rule approach~\cite{jko,joh2}.  We have calculated the
masses of positive and negative parity baryons ($B_{+}$ and $B_{-}$)
in the flavor octet and singlet representations.  We have found that
the quark condensates, which break chiral symmetry, induce the mass
splitting between $B_{+}$ and $B_{-}$.  When the chiral order
parameter $\langle \bar{q} q \rangle$ vanishes, we have found that
$B_{-}$ is degenerate with $B_{+}$ and also that they tend to become
massless.  This result may be contrasted with the observation made by
DeTar and Kunihiro.  We have also investigated the $\pi NN^{*}$
coupling in the same framework of the QCD sum rule~\cite{joh2}.
(Throughout this paper $N^*$ denotes a negative parity nucleon, e.g.\
$N(1535)$.)  There it is found that the $\pi NN^{*}$ coupling vanishes
in the chiral and soft-pion limit.  On the other hand, as opposed to
our finding, Kim and Lee have obtained a non-vanishing $\pi NN^{*}$
coupling when an alternative interpolating field that contains a
derivative is used for $N^{*}$~\cite{kimlee}.

The purpose of the present letter is to point out the importance of
chiral symmetry for properties of positive and negative parity
nucleons.  It turns out that when we treat two kinds of nucleons, there
are two ways to assign chiral transformations of baryons.  As a
consequence we can construct different chiral effective models based
on the two chiral realizations, and then we clarify why different
results were obtained from various approaches discussed above.  We
study masses, $\pi NN^*$ couplings and axial charges of $N$ and
$N^{*}$.  We also discuss the role of the chiral symmetry breaking in
the $N$-$N^*$ mass splitting.  To accomplish our purpose, we first
consider chiral transformations of $N$ and $N^{*}$, and then
by using the linear sigma model, we discuss the behaviors of
$N$ and $N^*$ under the chiral phase transition.

Let us start with considering the chiral transformation for one
nucleon $N$.  To be definite we will consider the $SU(2)_R \times
SU(2)_L$ chiral group.
We assume
that the nucleons belong to linear representations of the chiral group.
This may not be the case when chiral symmetry is spontaneously broken,
where
nonlinear representations are also possible.
In that case, however, it is difficult to
study the transition properties of the chiral symmetry.
Therefore, we will adopt linear representations for a consistent
study of chiral restoration properties.

The chiral transformation for $N$ is defined by
\begin{equation}
   N_R \longrightarrow R N_R \ , \hspace{1cm} N_L \longrightarrow L N_L \ ,
   \label{chitra1}
\end{equation}
where $R$ ($L$) is an element of $SU(2)_{R}$ ($SU(2)_{L}$), and
$N_{R}$ ($N_{L}$) are the right (left) component of the Dirac spinors,
satisfying $\gamma_{5} N_{R} = N_{R}$ and $\gamma_{5} N_{L} = -
N_{L}$.
Eq.(\ref{chitra1}) is no more than  definition;
the transformation for the ``right'' (``left'') handed nucleon is
just called the  ``right'' (``left'')  transformation.

Chiral symmetry does not allow a mass term for $N$ in the Lagrangian, 
so that $N$ is a massless particle in the Wigner phase (the chiral 
restored phase) and the nucleon mass is generated by spontaneous 
chiral symmetry breaking.  To find the chiral partner of $N$, we 
consider the commutation relation of the generators of the axial 
transformation $X^{a}(a = 1,2,3)$ and $N$ in the Wigner phase :
\begin{equation}
  [ X^a , N ] = - i \gamma_5 {\tau^a \over 2} N \ ,
    \label{comq5N}
\end{equation}
where $\tau^a$ is the isospin matrix.  This commutation relation
follows from the chiral transformation (\ref{chitra1}).
The commutation relation (~\ref{comq5N}) implies that
the chiral partner of $N=N_R + N_L$ is $\gamma_5 N = N_R - N_L$ and
that no additional particle is necessary to complete the chiral
multiplet $(\hal{1},0) \oplus (0,\hal{1})$.  It should be noted that
because $N$ is massless in the Wigner phase, we may regard $N_R$ and
$N_L$ as independent particles.  In other words a massive fermion in
NG phase (the chiral broken phase), which has four components, splits
into two massless fermions with two components each in the Wigner
phase and they form the pair of the chiral multiplet.

Next, we consider additional nucleon $N^{*}$ with negative parity.
First we introduce two nucleon fields $N_{1}$ and $N_{2}$ each of
which belongs to the chiral multiplet $(\hal{1},0) \oplus
(0,\hal{1})$.  The physical nucleons $N$ and $N^{*}$ are linear
combinations of $N_{1}$ and $N_{2}$, when the Lagrangian has the
mixing terms.  As anticipated, there are two possible assignments of
chiral transformations.  In the first scheme, which we call the
``naive assignment'', both $N_1$ and $N_{2}$ transform in the same
way.  In the second scheme, which we call the ``mirror assignment'',
the second nucleon transforms in the reversed way to the first
nucleon.

In the naive assignment the chiral transformation for $N_{1}$ and
$N_{2}$ is defined by
\begin{eqnarray}
  N_{1R} \longrightarrow R N_{1R}\  & , \hspace{1cm}&  N_{1L}
  \longrightarrow L N_{1L} \ , \label{naidef1} \\
  N_{2R} \longrightarrow R N_{2R}\  & , \hspace{1cm}&  N_{2L}
  \longrightarrow L N_{2L} \ . \label{naidef2}
\end{eqnarray}
Chiral symmetry requires again that these two nucleons must
be massless in the Wigner phase.
In order to find
their chiral partners, we consider the commutation
relations which follow from (\ref{naidef1}) and  (\ref{naidef2}):
\begin{equation}
   [X^{a}, N_{1}]  = - \hal{1} i\gamma_{5} \tau^{a} N_{1}
   \ , \hspace{1cm}
   [X^{a}, N_{2}]  = -\hal{1} i\gamma_{5} \tau^{a}
   N_{2} \ . \label{comQ5nai}
\end{equation}
These relations show that $N_{1}$ and $N_{2}$ belong to
the multiplet
$(\hal{1},0) \oplus (0,\hal{1})$
separately.
Therefore, as in the previous case, the chiral
partner of $N_{1}$ is $\gamma_{5} N_{1}$, and that of $N_{2}$ is
$\gamma_{5}N_{2}$. This situation does not change even if many
nucleons with the naive transformation are considered.

We note that
the commutation relations (~\ref{comQ5nai}) imply that
$g_{A} = 1$ for both $N$ and $N^*$
in the linear sigma model unless we
introduce derivative couplings of $\sigma$ and $\pi$ with the nucleon.  It
is important to note that the sign of the axial charge $g_{A}$ of
$N^{*}$ is the same as the one of $N$.  As we will see later, in the
mirror case the axial charge of $N^{*}$ has the opposite sign to $N$.
This is notable difference between the naive model and the mirror model.

In the mirror assignment, denoting the two nucleon fields by
$\psi_{1}$ and $\psi_{2}$, the transformation rule is defined as
\begin{eqnarray}
   \psi_{1R} \longrightarrow R \psi_{1R} \  & , \hspace{1cm} &   \psi_{1L}
   \longrightarrow L \psi_{1L} \ , \label{mirdef1} \\
   \psi_{2R} \longrightarrow L \psi_{2R} \  & , \hspace{1cm} &   \psi_{2L}
   \longrightarrow R \psi_{2L} \ . \label{mirdef2}
\end{eqnarray}
The right (left) component of $\psi_2$ transforms as the left (right)
component of $\psi_1$.  The reason that this assignment is possible is
that the left- and right-handedness of the fermion, which is
determined by the eigenvalue of $\gamma_{5}$, is independent of that
of chiral symmetry, although we use the same terminology.  The
chirality of fermion specifies representations of the Lorentz group
for the fermion while the chirality of chiral symmetry is associated
with the internal chiral symmetry $SU(2)_{R} \times SU(2)_{L}$.
In this case, we are allowed to introduce a chirally invariant mass
term
\begin{equation}
     m_{0}( \bar{\psi_2} \gamma_{5} \psi_1 - \bar{\psi_1}
      \gamma_{5} \psi_2 ) = m_0 (\bar{\psi_{2L}} \psi_{1R} -
        \bar{\psi_{2R}} \psi_{1L} - \bar{\psi_{1L}} \psi_{2R} +
        \bar{\psi_{1R}} \psi_{2L}) \ . \label{chinvmass}
\end{equation}
Therefore, the nucleons can have a finite mass $m_{0}$ when the chiral
symmetry is restored.

Using eqs. (\ref{mirdef1}) and (\ref{mirdef2}),
we find the following commutation relations
\begin{equation}
   [X^{a}, \psi_{1}]
   =  - i \gamma_5 \hal{\tau^{a}} \psi_{1} \ , \hspace{1cm}
   [X^{a}, \psi_{2}]
   =  + i \gamma_5 \hal{\tau^{a}} \psi_{2} \ .
\end{equation}
Note that the sign on the r.h.s.\ for $\psi_2$ is opposite to that for
$\psi_1$.  In order to obtain the physical nucleons $\psi_+$ and
$\psi_-$, we have to diagonalize the mass term (\ref{chinvmass})
because it is off-diagonal in the basis $(\psi_1, \psi_2)$.  In the
diagonalized basis, the commutation relations are given
by
\begin{equation}
   [X^{a}, \psi_{+}]  =  - \hal{\tau^{a}} \psi_{-} \ , \hspace{1cm}
   [X^{a}, \psi_{-}] =  - \hal{\tau^{a}} \psi_{+} \ ,
\label{comQ5mir}
\end{equation}
where $\psi_+ = {1 \over \sqrt{2}} ( \psi_1 + \gamma_5 \psi_2)$ and
$\psi_- = {1 \over \sqrt{2}} ( \gamma_{5} \psi_1 - \psi_2)$ in the
Wigner phase.  From the commutation relation (~\ref{comQ5mir}) we see
that $\psi_{+}$ and $\psi_{-}$ are transformed into each other under the
chiral transformation, and therefore, $\psi_{+}$ and $\psi_{-}$ belong
to the same multiplet of $SU(2)_{R} \times SU(2)_{L}$.  In this way,
$\psi_+$ and $\psi_-$ are considered to be chiral partners of each
other.

More explicitly, in the group theoretical language, $\psi_{1R}$ and
$\psi_{2L}$ belong to $(\hal{1},0)$ and $\psi_{1L}$ and $\psi_{2R}$
belong to $(0,\hal{1})$.  Because the nucleons have masses in the
Wigner phase, we need four components to represent each nucleon.
Thus it is appropriate to introduce
\begin{eqnarray}
   \Psi_r & \equiv & {1 \over \sqrt{2}} (\psi_{1R} \oplus \psi_{2L}) \\
   \Psi_l & \equiv & {1 \over \sqrt{2}} (\psi_{2R} \oplus \psi_{1L}) \ ,
\end{eqnarray}
which have four independent components each.  We see that $\Psi_r$
belongs to $(\hal{1},0)$ and $\Psi_l$ belongs to $(0,\hal{1})$ and
that the parity eigenstates $\psi_+$ and $\psi_-$ are given by linear
combinations of $\Psi_r$ and $\Psi_l$.  Thus $\psi_+$ and $\psi_-$
belong to the same multiplet $(\hal{1},0) \oplus (0,\hal{1})$ and the
chiral partner of $\psi_+$ is $\psi_-$.

In order to see the
above argument more concretely, we now construct
linear sigma models for the two kinds of nucleons which satisfy the
above chiral symmetries.

\begin{figure}
\epsfxsize = 15cm
\epsfbox{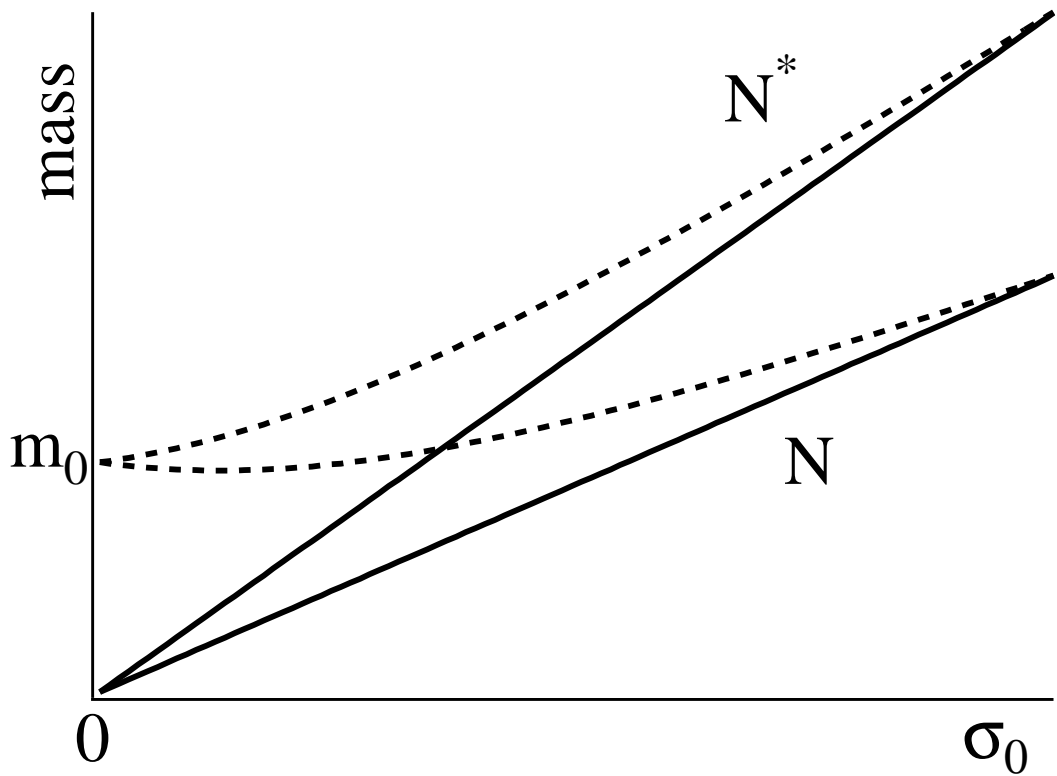}
\caption{A schematic plot of $\sigma_0$ dependences of
         $N$ and $N^*$ masses in the naive model (the solid line)
         and the mirror model (the dashed line).\label{fig}}
\end{figure}

First we consider the case of the naive assignment.  Considering the
transformation rule for the meson field $ M \equiv \sigma + i
\vec{\tau} \cdot \vec{\pi} \rightarrow L M R^{\dagger}$, we can easily
write down a renomalizable chiral invariant Lagrangian:
\begin{eqnarray}
   {\cal L}_{\rm{nai}} & = &\bar{N_1} \delslash N_1 + \bar{N_2}
\delslash
   N_2 + a \bar{N_1} (\sigma + i \gamma_5 \vec{\tau} \cdot \vec{\pi})
   N_1 + b \bar{N_2} (\sigma + i \gamma_5 \vec{\tau} \cdot \vec{\pi})
   N_2 \nonumber \\
         & & + c \{ \bar{N_2} (\gamma_5 \sigma + i \vec{\tau} \cdot
         \vec{\pi}) N_1  - \bar{N_1} (\gamma_5 \sigma + i \vec{\tau} \cdot
         \vec{\pi}) N_2 \} + {\cal L}_{M}  \, , \label{ordsu2lag}
\end{eqnarray}
where $a$, $b$ and $c$ are coupling constants.
Here ${\cal L}_M$ is a chiral invariant meson Lagrangian which is not
important in the following discussion.  The fifth term in this
Lagrangian gives a mixing between $N_{1}$ and $N_{2}$.  The chiral
symmetry breaks down spontaneously with a finite vacuum expectation
value of the sigma meson, $\sigma_{0} \equiv \bra{0} \sigma \ket{0}$.  To
obtain physical nucleons $N_{+}$ and $N_{-}$ in the NG phase, we have
to diagonalize the mass matrix $M$ which is given by
\begin{equation}
  M \sim \sigma_{0} \left(
     \begin{array}{cc}
        a & -\gamma_{5} c  \\
        \gamma_{5} c & b
      \end{array}
    \right)
\end{equation}
This matrix can be diagonalized by the physical nucleon fields
\begin{equation}
   \left(
       \begin{array}{c}
        N_{+}  \\
        N_{-}
       \end{array}
   \right) = {1 \over \sqrt{2 \cosh \delta }}
        \left(
             \begin{array}{cc}
             e^{\delta/2}      & \gamma_{5} e^{-\delta/2}  \\
                \gamma_{5} e^{-\delta/2} & - e^{\delta/2}
             \end{array}
        \right)
        \left(
             \begin{array}{c}
                N_{1}  \\
                N_{2}
             \end{array}
        \right) \label{N+N-}
\end{equation}
where the mixing angle $\delta$ is defined by $\sinh \delta = -(a + b)/
2 c$.  In this basis the masses of $N_{+}$ and $N_{-}$ are given by
\begin{equation}
   m_{\pm} = {\sigma_{0} \over 2} \left( \sqrt{(a + b)^{2} + 4c^{2}}
   \mp (a-b) \right) \ . \label{naimass}
\end{equation}
We present a schematic plot of $m_{\pm}$ as functions of
$\sigma_0$ in Fig.\ref{fig}.  In the Wigner phase, i.e.\@ when
$\sigma_{0} \rightarrow 0$,
both $N_+$ and $N_-$ become massless and get degenerate.
However, this degeneracy is trivial rather than due to
chiral symmetry,
because all the nucleons with the naive assignment are massless in the
Wigner phase. Similarly, the mass difference of $N_{+}$ and $N_{-}$
is caused by the choice of the coupling parameters, $a$ and $b$ and
therefore it is independent of chiral symmetry.

It should be noted that the physical nucleon fields $N_{+}$ and
$N_{-}$ decouple from each other after the diagonalization, when we
truncate the meson-nucleon coupling Lagrangian at the non-derivative
Yukawa term, because the coupling term of (\ref{ordsu2lag}) is
factored out by the mass matrix $M$.  This implies that the
off-diagonal Yukawa coupling $g_{\pi N_{+}N_{-}}$ vanishes in the soft
pion limit, when all the derivative couplings are neglected.  This
result is qualitatively consistent with the observed $g_{\pi NN(1535)}
\sim 1$ which is strongly suppressed in comparison with $g_{\pi NN}
\sim 13$.  In fact, up to the nonderivative Yukawa coupling, this
sigma model is reduced to the sum of two independent sigma model even
without the $\sigma$ condensation, since the mixing angle $\delta$ is
independent of $\sigma_{0}$.  Namely the parameter $c$, which appears
in the off-diagonal Yukawa coupling, is a superficial parameter.

Next we turn to the discussion of the mirror model.  In this model the
negative parity nucleon is assumed to follow the second chiral
transformation as given in (\ref{mirdef2}).  The renomalizable chiral invariant
Lagrangian is then
\begin{eqnarray}
     {\cal L}_{\rm{DK}} & = & \bar{\psi_1} i \delslash \psi_1 +
           \bar{\psi_2} i \delslash \psi_2
        + m_{0}( \bar{\psi_2} \gamma_{5} \psi_1 - \bar{\psi_1}
            \gamma_{5} \psi_2  )
                \nonumber \\
      & &    + a \bar{\psi_{1}} (\sigma + i \gamma_5 \vec{\tau} \cdot
           \vec{\pi}) \psi_{1} +
        b \bar{\psi_{2}} (\sigma - i \gamma_5 \vec{\tau} \cdot
        \vec{\pi}) \psi_{2} \nonumber \\
      & & + {\cal L}_{M} \ ,
  \label{mirsu2lag}
\end{eqnarray}
where $a = g_{2} - g_{1}$ and $ b = - g_{1} - g_{2}$ using the
parameters
in Refs.~\cite{dk,nemoto}.  This Lagrangian was first proposed and
studied
by DeTar and Kunihiro~\cite{dk}.

In the same way as in the naive case, with the spontaneously chiral
symmetry breakdown, we diagonalize the mass matrix by
$\psi_{+}$ and $\psi_{-}$:
\begin{equation} \label{mirbasis}
  \left( \begin{array}{c} \psi_+ \\ \psi_- \end{array} \right) =
  \frac{1}{\sqrt{2 \cosh \delta}} \left(
             \begin{array}{cc}
             e^{\delta/2}      & \gamma_{5} e^{-\delta/2}  \\
                \gamma_{5} e^{-\delta/2} & -e^{\delta/2}
             \end{array}
        \right)
    \left( \begin{array}{c} \psi_{1} \\ \psi_{2} \end{array} \right)
\end{equation}
with $\sinh \delta = -(a + b) \sigma_0 / 2 m_0$. Note that the mixing
angle is dependent on the chiral order parameter $\sigma_0$.
In the basis (~\ref{mirbasis}) the masses of $\psi_+$ and $\psi_-$ are
given by
\begin{equation}
  m_\pm = {1 \over 2} ( \sqrt{ (a+b)^2 \sigma_0^2 + 4 m_0^2}
              \mp (a -b) \sigma_0 ) \ , \label{mirmass}
\end{equation}
A schematic plot of $m_{\pm}$ as functions of  $\sigma_0$
is presented in Fig.\ref{fig}.
In the Wigner phase these nucleons are degenerate with a finite mass 
$m_0$.  This shows the nucleon masses are generated by $m_0$, which is 
quite different from the mechanism of mass generation in the naive 
model, where the chiral symmetry breaking creates the nucleon masses.  
We see that the mass splitting between $\psi_+$ and $\psi_-$ is caused 
by the spontaneously chiral symmetry breaking.  In this sense $m_0$ is 
the most important parameter in the mirror model.  The case $m_0 = 0$ 
is special, because we can not distinguish the naive model and the 
mirror model.

We note two more differences in the mirror model from the naive model.  
First, meson couplings between $\psi_+$ and $\psi_-$ no longer vanish 
unlike the naive case and can remain finite, because the coupling 
matrix differs from the mass matrix and need not be diagonalized in 
the basis $(\psi_{+}, \psi_{-})$ that diagonalizes the mass matrix.  
Second, the commutation relations between the axial charges 
$Q_{5}^{a}$ and the 
nucleon fields are quite different,
\begin{eqnarray}
   [Q_{5}^{a}, \psi_{+}] & = &  \hal{\tau^{a}}(\tanh\delta \,
     \gamma_{5} \psi_{+} + {1 \over \cosh\delta} \psi_{-})
\label{mirq5com} \\
   \ [Q_{5}^{a}, \psi_{-}] & = & \hal{\tau^{a}}(-\tanh\delta \,
     \gamma_{5} \psi_{-} + {1 \over \cosh\delta} \psi_{+})
\end{eqnarray}
They imply that the axial charges are now given in the form of a 2
$\times$ 2 matrix and that the sign of $g_{A}^{\psi_{-}\psi_{-}}$ is
opposite to $g_{A}^{\psi_{+}\psi_{+}}$.
It would be of great interest to see the relative sign of
the axial charges experimentally, as it provides the key information on
the chiral structure of negative parity nucleon.  We summarize a
comparison between the naive and mirror model in Table ~\ref{sum}.

Finally, we comment on the QCD sum rule analysis on the negative
parity nucleon.
It turns out that the chiral
assignment in our QCD sum rule\cite{jko,joh2} corresponds to the naive
assignment.  In this analysis we have introduced $N$ and $N^{*}$ as
\begin{eqnarray}
     \bra{0} J(x) \ket{N} & = & \lambda_{N} u_{N}(x) \ ,
         \label{constN} \\
     \bra{0} J(x) \ket{N^{*}} & = & i \gamma_{5} \lambda_{N^{*}}
     u_{N^{*}}(x) \ .
         \label{constN*}
\end{eqnarray}
Then the same chiral transformation for $N^{*}$
as for $N$ follows.
This is the basis of our previous QCD sum rule results,  i.e.,
that $N$ and $N^{*}$ tend to become
massless as $\langle \bar{q}q \rangle \rightarrow 0$ and that the $\pi
NN^{*}$ coupling vanishes in the soft and chiral limit.

On the other hand, Kim and Lee found quite different results from
ours\cite{kimlee}.
While they have
used the same type of interpolating field for $N$ as our choice,
for $N^{*}$
they have adopted an alternative interpolating field $\eta_{N^{*}}$
that contains a derivative~\cite{leekim}.  Although the $\eta_{N^{*}}$ itself
transforms in the same way as $N$, the chiral structure of
$N^{*}$ is changed by the coupling of $N^{*}$ to $\eta_{N^{*}}$:
\begin{equation}
\label{MEkimlee}
        \langle 0 | \eta_{N^*}|N^*\rangle = i \lambda_{N^*} \gamma_5
z_\mu
        \gamma^\mu u_{N^*} \, ,
\end{equation}
where $z_\mu$ is an auxiliary space-like vector which is orthogonal to
the four momentum carried by the resonance state.  The $\gamma_{\mu}$
matrix on the r.h.s.\ of (\ref{MEkimlee}) changes the chirality and
thus makes $N^{*}$ being a mirror of $N$.
This is the reason why they have obtained a finite $\pi NN^{*}$
coupling.

We need further investigations on chiral properties of baryons, 
because we know very little which of the two assignments is realized 
in the physical nucleons.  As far as the sum rule analysis are 
compared, there is no strong preference of one of the two assignments.  
DeTar and Kunihiro reproduced the masses of $N$ and $N(1535)$ and the 
observed $g_{\pi NN(1535)}$ by choosing $m_0 = 270$ MeV \cite{dk}.  
The DK model and our extended model to $SU(3)$~\cite{nemoto} suggest 
that the suppression of the $\pi NN^*$ coupling is caused by the 
smallness of $m_0$, which is the key parameter characterizing this 
model.  For small $m_{0}$, the mirror assignment may give similar 
predictions for the masses of $N$ and $N^{*}$ as well as the $\pi 
NN^{*}$ coupling strengths and therefore can hardly be distinguished 
phenomenologically.  However, a notable difference between the two 
choices is the sign of the axial charge of $N^{*}$, 
$g_{A}^{N^{*}N^{*}}$.  It would be extremely interesting if the axial 
charge of $N^{*}$ is observed experimentally.

In summary, we have investigated the properties of the negative parity 
nucleon $N^*$ from the viewpoint of chiral symmetry.  We have two ways 
of assignments of chiral transformation for $N^{*}$, so that we obtain 
two linear sigma models based on them.  We have observed several 
qualitative differences of the properties of $N$ and $N^*$ between the 
two, which are summarized in Table ~\ref{sum}.  The origin of the 
differences is the chiral structures of the nucleons.  In the naive 
case, $N^{*}$ has nothing to do with $N$ in the sense that the $N^{*}$ 
belongs to different multiplets from the one of $N$, while in the 
mirror case $N^{*}$ belongs to the same multiplet as that of $N$, so 
that $N^{*}$ can be interpreted as the chiral partner of $N$.  If in 
the real world the mirror case is realized, we need to identify the 
chiral partner of $N(939)$ with the negative parity nucleon such as 
$N(1535)$ or $N(1650)$.

This work is supported in part by the Grant-in-Aid for scientific
research (A)(1) 08304024 and (c)(2) 08640356 and also by the
Grant-in-Aid for Encouragements of Young Scientists
of the Ministry of Education, Science and Culture of Japan.
The work of D.J. is supported by Research Fellowships of the 
Japan Society for the Promotion of Science for Young Scientists.

\begin{table}
\caption{\label{sum}}
\begin{tabular}{|c|c|c|}
      & naive assignment & mirror assignment\\
\hline
definition & $ N_{2R} \rightarrow R  N_{2R} \ , \
               N_{2L} \rightarrow L N_{2L} $ &
             $ \psi_{2R} \rightarrow L \psi_{2R} \ , \
               \psi_{2L} \rightarrow R \psi_{2L} $ \\
mass in the Wigner phase& 0 & $m_0$ (finite) \\
$\pi NN^* $ coupling & 0 & $ (a + b)/\cosh \delta$ \\
chiral partner & $N_+ \leftrightarrow \gamma_5 N_+ \ , \
                  N_- \leftrightarrow \gamma_5 N_- $ &
       $\psi_+ \leftrightarrow \psi_- $ \\
$g_{A}^{NN} g_{A}^{N^* N^*} $ & positive & negative  \\
role of $\sigma_0$ & mass generation & mass splitting \\
\end{tabular}
\end{table}

\begin{thebibliography}{9}
    \bibitem{hatsuda} T. Hatsuda and T. Kunihiro,
    \Journal{Phys. Rept.}{247}{221}{1994}.

    \bibitem{lee} B. W. Lee, {\it Chiral Dynamics}, Gordon and Breach,
   New York, (1972)

    \bibitem{coji} T.D. Cohen and X. Ji, \Journal{\PRD}{55}{6870}{1997}

    \bibitem{dk} C. DeTar and T. Kunihiro,
    \Journal{\PRD}{39}{2805}{1989}.

    \bibitem{lattice} C. E. DeTar and J. B. Kogut,
    \Journal{\PRL}{59}{339}{1987}; \Journal{\PRD}{36}{2828}{1987};
    S. Gottlieb, W. Liu, D. Toussaint, R. L. Renkin, and R. L. Sugar,
    \Journal{\PRL}{59}{1881}{1987}.
    
    \bibitem{inst} T. Sch\"afer and E.V. Shuryak, 
    \Journal{\PLB}{356}{147}{1995}.

    \bibitem{jko} D. Jido, N. Kodama and M. Oka,
    \Journal{\PRD}{54}{4532}{1996};
    D. Jido and M. Oka, {\it QCD sum rule for
    ${1 \over 2}^{-}$ Baryons}, TIT-preprint, hep-ph/9611322.

    \bibitem{joh2} D. Jido, M. Oka and A. Hosaka,
    \Journal{\PRL}{80}{448}{1998}

    \bibitem{kimlee} H. Kim and S.-H. Lee,
    \Journal{\PRD}{56}{4278}{1997}. 


    \bibitem{nemoto}  Y. Nemoto, D. Jido, M. Oka and A. Hosaka,
    \Journal{\PRD}{57}{4124}{1998}
    
    \bibitem{leekim} S.-H. Lee and H. Kim, 
    \Journal{\NPA}{612}{418}{1997}

\end{thebibliography}
\end{document}